# TWO RECENT kHz OUTER HELIOSPHERIC RADIO EMISSIONS SEEN AT VOYAGER 1 - WHAT ARE THE INTERPLANETARY EVENTS THAT TRIGGER THEM AND WHERE ARE THESE EVENTS WHEN THE RADIO EMISSIONS START?

W.R. Webber[1] and D.S. Intriligator[2]

1. New Mexico State University, Department of Astronomy, Las Cruces, NM , 88003, USA
2. Carmel Research Center, Space Plasma Laboratory, Santa Monica, CA 90406, USA

## ABSTRACT

We have examined instigating events at the Sun that may be responsible for two of the most recent outer heliospheric kHz emissions detected by the University of Iowa plasma wave detector on Voyager 1 starting at 2004.64 and 2006.39, respectively. These interplanetary events have been followed outward from the Sun using plasma and cosmic ray data from Ulysses and Voyagers 1 and 2. For both intervals of kHz emissions, events originating near the Sun that turn out to be the most intense events in this 11-year solar cycle as observed by the plasma and cosmic ray variations in the outer heliosphere, reach V1 and V2 which are near the heliospheric termination shock at almost the same time that the two kHz radio emissions turn-on. These two events which originate near the Sun about 2003.89 (the 2003 Halloween event) and at 2005.71, are also unusual in that they develop solar wind ram pressure waves in the outer heliosphere with maximum pressures ~6-8 times the average solar wind pressure when they reach the heliospheric termination shock. We believe that this study suggests a possible new paradigm for the origin of these latest kHz emissions. This new paradigm is different from the current one describing earlier kHz events. It is related to the strength of these pressure waves in the outer heliosphere and involves the arrival of these large pressure waves at the heliospheric termination shock rather than the arrival of a shock at the heliopause as is the current paradigm.

## Introduction

The kHz radio emissions detected by the University of Iowa plasma wave detectors on the Voyager spacecraft have provided a remarkable insight into conditions in the outermost heliosphere. The present paradigm is that these emissions originate when a large shock reaches the outermost boundary of the heliosphere, the heliopause (HP). This interaction then triggers plasma waves in the local interstellar (IS) medium at frequencies just above the local plasma frequency (Gurnett, 1993). To date perhaps 6-10 of these radio events have been observed over three solar 11-year cycles. The first kHz emission starting at 1983.84, and another starting at 1992.75 are by far the largest and best documented. The particular shocks and interplanetary disturbances that are believed to be the source of these kHz emissions have their origin near the Sun. For the two earlier kHz emissions noted above, using an identification near the Sun based on large cosmic ray decreases at the Earth, the travel times to the onset of the kHz emission has been determined to be ~415±5 days. These times along with estimates of the speed profile of the shocks with distance from the Sun have allowed estimates of the HP distance in the range 120-160 AU and have served to define the scale size of the heliosphere (Gurnett, et al., 1993).

In the most recent solar cycle #23 starting at 1997.5, there have been no really large kHz radio emissions like those in cycles 21 and 22, but there have been several smaller ones (Gurnett, Kurth and Stone, 2003; Gurnett, 2007, see Figure 1). Voyagers 1 and 2 (V1 and V2) are now at much larger distances from the Sun; at 2005.0 these distances are ~95 and 76 AU, respectively. As a result they may be used to better determine the characteristics of these interplanetary (IP) shocks at large distances from the Sun as they pass V1 and V2, in some cases near to or beyond the heliospheric termination shock (HTS). From the timing of these shocks as they reach V1 and V2 one is also able to greatly reduce the timing uncertainties as these shocks propagate to the HTS and beyond.

In this paper we examine two recent kHz emissions reported by the Iowa group. The first had its onset at about 2004.64. The second had its onset at about 2006.39. These emissions are illustrated in Figure 1, provided to us by Gurnett, 2007. Neither of these radio events has been discussed in any detail in the literature although the onset times for the kHz emission have been presented in figures discussed by Gurnett, 2007.

**The Heliospheric kHz Emissions with an Onset at 2004.64**

These kHz emissions have a possible origin near the Sun at the time of the well documented series of events collectively known as the "Halloween" 2003 events (Intriligator, et al., 2005) occurring at about 2003.83 at the Earth. As Gurnett, 2007, note, however, this timing would lead to a total travel time from the Earth to the onset of the kHz radiation of only 0.81 years, much shorter than the travel time of ~1.15 years for the earlier large kHz events in solar cycles 21 and 22.

Let us consider the implications of the "Halloween" 2003 event at the Earth as the possible instigating event for these kHz emissions. We have an onset time ~2003.83 at the Earth at which time an exceptionally strong shock was observed and also a Forbush decrease ~26% was seen by the Climax NM. Later at 2003.87 a solar wind speed jump of ~200 km$\cdot$s$^{-1}$ to >850 km$\cdot$s$^{-1}$ was observed in the plasma detector on Ulysses then at ~5 AU and ~4° N. At V2, then at 73.4 AU, a solar wind speed jump of ~100 km$\cdot$s$^{-1}$ to ~560 km$\cdot$s$^{-1}$ was observed in the plasma detector at 2004.33 (event #7 in Figure 2). This speed jump at V2 was considered to be due to the passage of the Halloween event beyond V2 (Richardson, et al., 2005). An increase of a factor ~4 was seen in 2-3 MeV protons peaking at 2004.31 and a decrease of >70 MeV particles ≥ 10% was also observed by CRS detectors on Voyager 2. These intensity changes have also been attributed to the arrival of the Halloween event at V2 (Webber, et al., 2007). This event, which was one of the largest in solar cycle 23 by all measures, near the Earth, at Ulysses and also in the outer heliosphere at V2, took ~0.50 years to travel from the Earth to V2 implying an average speed ~665 km$\cdot$s$^{-1}$ (Intriligator, et al., 2005).

At V1 there is a decrease in the >70 MeV rate ~6% at 2004.59, followed by a further decrease ~5% starting at 2004.71. Collectively these decreases mark the arrival of this same shock at V1, then at 92.3 AU and just inside the HTS at 94 AU (Webber, et al., 2007).

The time period for the arrival of the "Halloween" shock at V1 just inside the HTS therefore brackets the time of onset of the kHz radio emission at 2004.64. So, in effect, the onset of the kHz radio emission for this event coincides very closely with the arrival of the "Halloween" shock at the HTS and at V1 just 2 AU inside the HTS.

**The Heliospheric kHz Radio Emission with an Onset at 2006.39**

A potential candidate for the origin of this kHz radio event is the intense solar activity observed in September, 2005. In September, 2005, there was a large Forbush decrease of cosmic rays of amplitude ~15% seen by the Climax NM, with an onset on September 10th (2005.70). This event was also observed at Ulysses, now at 4.8 AU and -28° S, (in latitudinal alignment with V2) on about September 15th as a large plasma speed jump ~280 $km \cdot s^{-1}$ to >700 $km \cdot s^{-1}$. At V2, at 79.0 U, the largest solar wind speed jump yet observed in the outer heliosphere in this solar cycle, ~140 $km \cdot s^{-1}$ to a maximum solar wind speed of ~520 $km \cdot s^{-1}$ was observed. This occurred at 2006.17 (Richardson, et al., 2006) (event #10 in Figure 2) and the >70 MeV cosmic ray rate decreased by ~14% at 2006.21 about 0.20 years before the onset of the kHz emission (Webber, et al., 2007; Intriligator, et al., 2005). Both the plasma and cosmic ray events at V2 have been ascribed to the September events at the Earth (Richardson, et al., 2006; Webber, et al., 2007). The travel time from the Earth to V2 of 0.48 year leads to an average speed ~760 $km \cdot s^{-1}$, the highest average speed observed for any event in the outer heliosphere. At V1 at 99.7 AU, and beyond the HTS, a decrease ~5-6% was seen in the >70 MeV rate by the CRS experiment at 2006.51 (Webber, et al., 2007) and is believed to mark the arrival of this event at V1.

The onset of the kHz radio emission at 2006.39 thus occurs between the arrival of this event at V2 at 79 AU inside the HTS and its arrival at V1 at 99.7 AU outside the HTS. This would again imply a turn-on associated with the shock arrival near the HTS, not the HP.

**Discussion: The Significance of Outward Propagating Pressure Waves**

The two kHz radio emission intervals discussed here serve to emphasize an under appreciated problem of first identifying the correct instigating event on the Sun and then following it through the heliosphere using Ulysses, V2 and V1 data when possible to determine the time when it reaches a particular point in the heliosphere, for example, the HTS or HP, at which time the kHz radio emission first turns on.

The real basis for understanding the onset of these kHz emissions thus becomes trying to understand what are the most important features of the instigating event that lead to the turn-on of the kHz radiation in the 1st place. In this connection we would like to note the significance of large pressure waves that develop during the propagation of the largest IP shocks through the heliosphere (Figure 3). This figure shows the daily average solar wind ram pressures measured at V2 between 2001 and the time it crossed the HTS at 83.7 AU at 2007.66. The two largest pressure waves at V2 in solar cycle 23 occur at the times that the Halloween event and the September 2005 event pass V2. These pressure waves, with peak pressures ~5-10 times the average solar wind pressure, are known to directly and immediately influence the HTS, its location and the structure of the heliosheath beyond the HTS, Webber, 2005; Richardson, et al., 2006 (see e.g., Washimi, et al., 2007, for an example of the effect of a pressure wave with ~1.5 times the average solar wind pressure impacting the HTS). Such large pressure waves impacting the HTS may provide the densities beyond the HTS required to produce the observed kHz emission.

**Summary and Conclusions**

In this paper we have examined the possibilities for the instigating events on the Sun that are responsible for two of the most recent outer heliospheric kHz radio emissions that began on 2004.64 and 2006.39 as observed by the Iowa plasma wave detector. In this study we utilize observations of the IP events as they pass V2 and V1 in the outer heliosphere between ~72 and 99 AU.

For both of the kHz emissions the particular instigating events that reach V1 and V2 which are near the HTS at the time of the turn-on the kHz radiation are the two largest in the outer heliosphere in solar cycle 23, in terms of solar wind speed jump (~ shock strength), galactic cosmic ray decreases and magnetic field increase. These two events are also particularly significant for their solar wind ram pressure waves that reach daily average pressures ~6-8 times the normal average solar wind pressure for times ~26 days or more. We plan to further investigate the passage of these instigating events from the Sun to the HTS and HP as well as earlier events using three-dimensional solar wind models (Intriligator, et al., 2008).

We, therefore, suggest that for these two most recent kHz emissions, strong arguments from both the timing and shock strength point of view can be made for the onset of the radio emission to be coincident with the arrival of the instigating interplanetary disturbances at the HTS rather than the HP. Theories for the origin of this kHz radiation in the Heliosheath (e.g., Zank, et al., 1994) need to be re-examined utilizing the much better current understanding of conditions there, following the latest V2 observations at the HTS and beyond. This study should include the effects of the large pressure waves that we identify in this paper and their encounter with the HTS, a topic only recently explored by Washimi, et al., 2007.

We note that McNutt, et al., 1988, originally suggested that the onset of the heliospheric radio emissions might be due to the arrival of solar wind disturbances at a hypothesized heliospheric termination shock. Subsequent attempts to explain the radio emissions in terms of large changes in density either in the foreshock region or in the region just beyond the HTS have had difficulties in explaining the radio emission at frequencies ≥ 2 kHz (e.g., Cairns and Zank, 2001) leading eventually to the current paradigm developed earlier by Gurnett (1993) and coworkers involving the heliopause and the presumed higher (electron) densities present there.

We should also note that there appear to be at least two types of radio emissions in the 2-5 kHz range. One type drifts significantly to higher frequency as a function of time, the other remains at a nearly constant frequency from2.0-2.6 kHz. The two events discussed here appear to be of the 1st type whereas, for example, the two earlier giant kHz events in 1981 and 1992, upon which the earlier theories were developed, are mainly the 2nd type showing little drift. It could be, for example, that the two different types of kHz events represent interactions of the outward moving solar disturbance near the HTS in one case and beyond the HP in the other.

In any case what the new theories should be is unclear. Certainly the timing relationships of earlier kHz radio emissions in solar cycles 21 and 22 need to be re-examined using the plasma and cosmic ray features of various possible instigating events including the development of these pressure pulses as the disturbance propagates outward in the heliosphere past V1 and V2. For example, the impact of these giant ram pressure pulses on the heliosheath beyond the HTS may lead to elevated heliosheath densities perhaps rivaling those found at the HP and beyond. Until such studies are completed it is not appropriate to claim that all kHz emissions observed by Voyager, including those observed earlier in solar cycles 21 and 22, have their origin near the HTS. It may be that shocks arriving at both locations can produce the observed kHz emission, depending on their specific characteristics and the details of their interaction with the HTS and HP, respectively.

Acknowledgments: The authors express their appreciation to the MIT plasma experimenters and the Iowa plasma wave experimenters on Voyager for making their data available on their web-sites and for valuable discussions. The work at the Carmel Research Center was supported by NASA grant NNX08AE40G and by the Carmel Research Center.

## References

Cairns, I.H. and G.P. Zank, Turn-on of radiation beyond the heliosphere and its propagation into the inner heliosphere, The Outer Heliosphere: The Next Frontier, Editors K. Scherer, et al., Pergamum, 2001

Gurnett, D.A., W.S. Kurth, S.C. Allendorf and R.L. Poynter, Radio emission from the heliopause triggered by an interplanetary shock, Science, 262, 199-203, 1993

Gurnett, D.A., Heliospheric radio emissions, Space Sci. Rev., 72, 243-254, 1995

Gurnett, D.A., W.S. Kurth and E.C. Stone, The return of the heliospheric 2-3 kHz radio emission during solar cycle 23, GRL, 30, Issue 23, pp. SSC 8-1, 2209, doi:10.1029/ 2003GL018514.

Gurnett, D.A., Private Communications to W.R. Webber and D.S. Intriligator, 2007, 2008

Intriligator, D.S., W. Sun, M. Dryer, C. Fry, C. Deehr, and J. Intriligator, From the Sun to the outer heliosphere: Modeling and analyses of the interplanetary propagation of the October/November (Halloween) 2003 solar events, J. Geophys. Res., 110, A09S10, doi:10.1029/2004JA010939, 2005.

Intriligator, D.S., W. Sun, A. Rees, T. Horbury, W.R. Webber, C. Deehr, T. Detman, M. Dryer, and J. Intriligator, Three-dimensional simulations of shock propagation in the heliosphere and beyond, in particle acceleration and transport in the heliosphere and beyond, 7th Annual International Astrophysical Conference, edited by G. Li, Q. Hu, O. Verkhoglyadova, G. Zank, R. Lin and J. Luhmann, AIP 10399, 375-383, 2008.

McNutt, J.R.L., A solar wind trigger for the outer heliospheric radio emissions and the distance to the terminal shock, Geophys. Res. Lett., 15, 1307-1310, 1988

Richardson, J.D., C. Wang, J.C. Kasper and Y. Liu, Propagation of the October/November, 2003 CME's through the heliosphere, Geophys. Res. Lett., 32, L03503, doi:10.1029/ 2004GL020679, 2005

Richardson, J.D., et al., Source and consequences of a large shock near 79 AU, Geophys. Res. Lett., 33, L23107, doi:10.1029/2006GL027983, 2006

Richardson, J.D., C. Wang and M. Zhang, Plasma in the outer Heliosphere and the Heliosheath, Physics of the Inner Heliosphere, CP 858, AIP, 110-115, 2006.

Washimi, H., G.P. Zank, Q. Hu, T. Tanaka and K. Munakata, A forecast of the heliospheric termination shock position by three-dimensional MHD simulations, ApJ, 670, L139-L142, 2007

Webber, W.R., An empirical estimate of the heliospheric termination shock location with time with application to the intensity increases of MeV protons seen at Voyager 1 in 2002-2005, J. Geophys. Res., 110, A10103, doi:10.1029/2005JA011209, 2005

Webber, W.R., A.C. Cummings, F.B. McDonald, E.C. Stone, B. Heikilla and N. Lal, Passage of a large interplanetary shock from the inner heliosphere to the heliospheric termination shock and beyond: Its effects on cosmic rays at Voyagers 1 and 2, GRL, 34, 20, L20107, doi: 10.1029/2007GL031339, 2007

Zank, G.P., I.H. Cairns, D.J. Donahue and W.H. Matthews, Radio emission and the heliospheric termination shock, J. Geophys. Res., 99, 14729-14735, 1994

## Figure Captions

**Figure 1:** kHz radio emissions observed by V1 (courtesy of D.A. Gurnett, 2007).

**Figure 2:** Daily average solar wind speed measured by the MIT plasma probe on V2. Numbered events indicate the passage of various shocks as discussed in the text.

**Figure 3:** Daily average solar wind pressure measured by MIT plasma probe on V2. Numbered events indicate the passage of various shocks as discussed in the text.

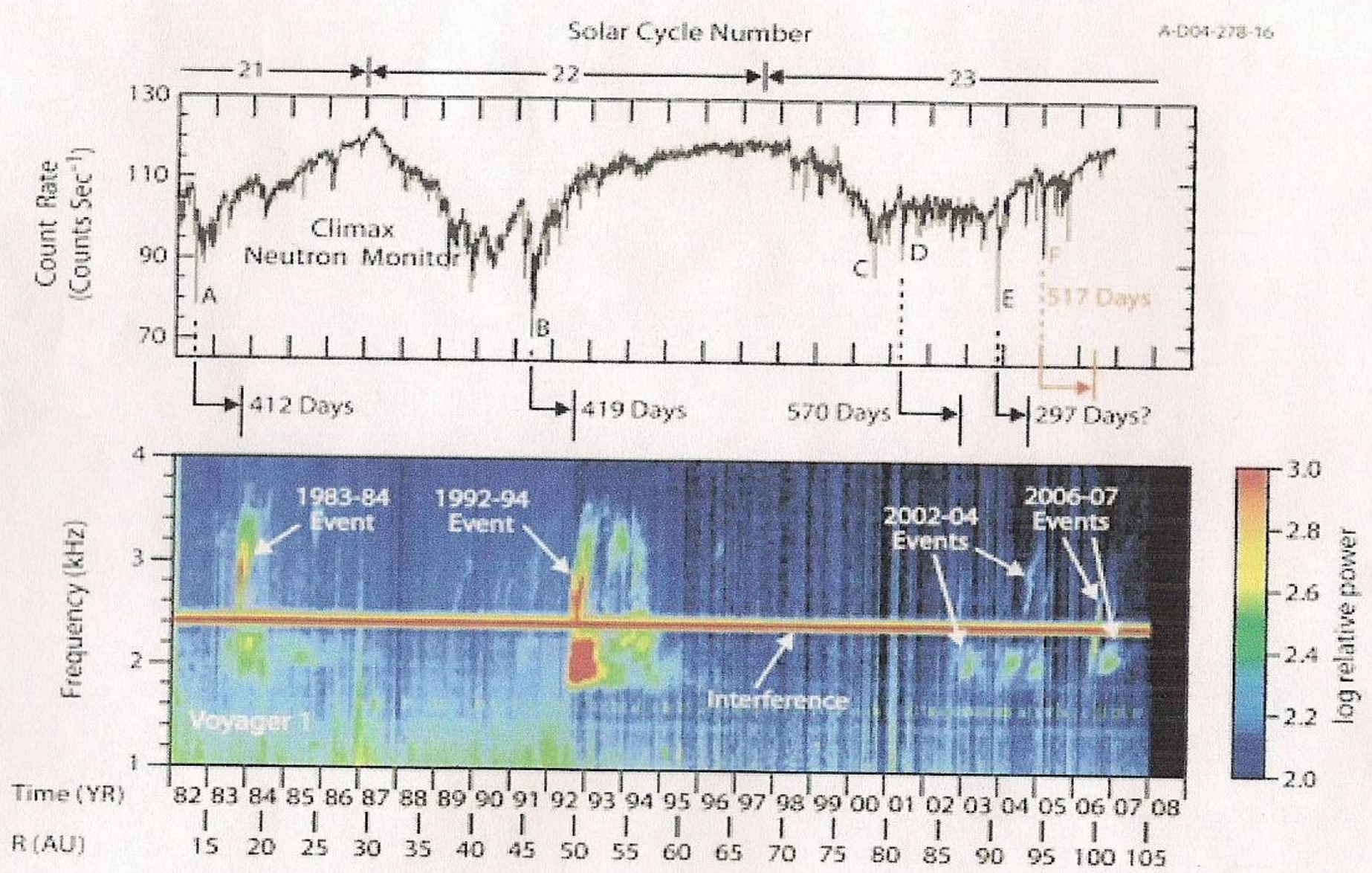
Solar Cycle Number
A-D04-278-16
21
22
23
Count Rate (Counts Sec$^{-1}$)
130
110
90
70
Climax Neutron Monitor
A
B
C
D
E
F
517 Days
412 Days
419 Days
570 Days
297 Days?
Frequency (kHz)
1983-84 Event
1992-94 Event
2002-04 Events
2006-07 Events
Interference
Voyager 1
log relative power
3.0
2.8
2.6
2.4
2.2
2.0
Time (YR)
82 83 84 85 86 87 88 89 90 91 92 93 94 95 96 97 98 99 00 01 02 03 04 05 06 07 08
R (AU)
15 20 25 30 35 40 45 50 55 60 65 70 75 80 85 90 95 100 105

FIGURE 1

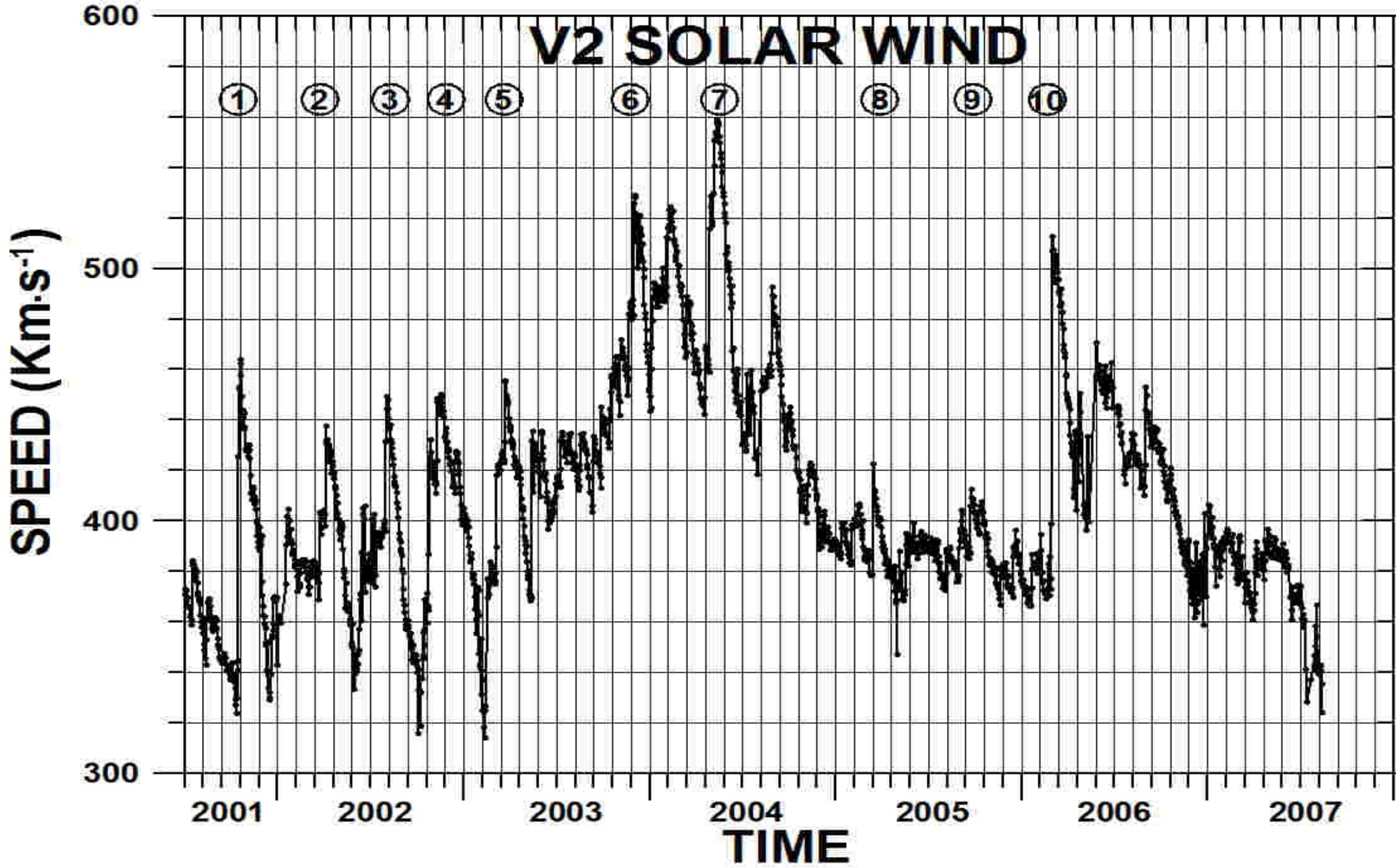

FIGURE 2

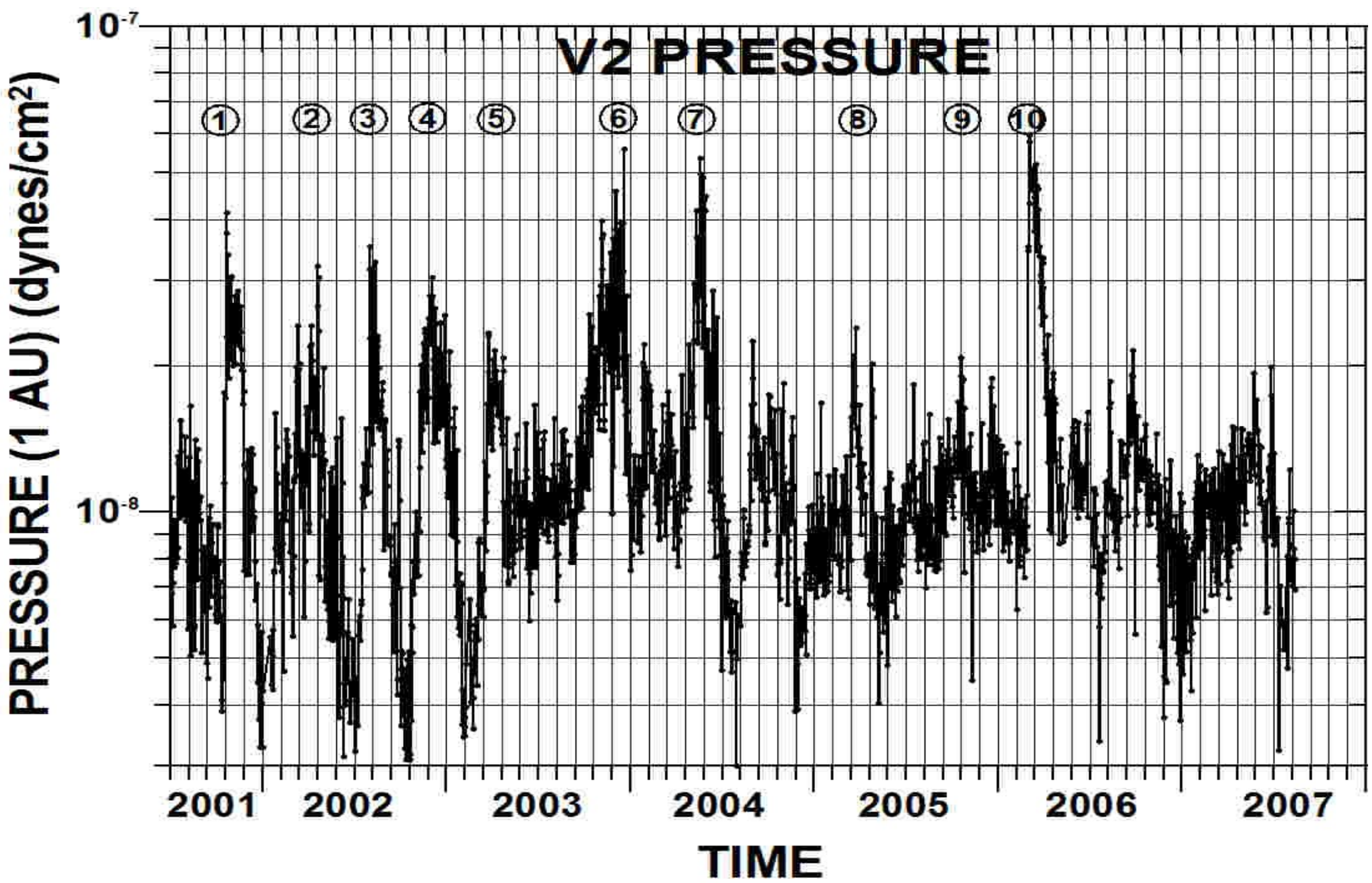

FIGURE 3